\documentclass[a4paper,fleqn,usenatbib, twocolumn]{mnras}

\usepackage{newtxtext,newtxmath}
\usepackage[T1]{fontenc}
\usepackage{ae,aecompl}
\usepackage{multirow}
\usepackage{graphicx}    
\usepackage{amsmath}    
\usepackage{amssymb}    
\usepackage{widetext}
\usepackage{ulem}
\usepackage{hyperref}
\usepackage[export]{adjustbox}

\def\araa{ARA\&A}
\def\apj{ApJ}
\def\apjs{ApJS}
 
\def\jcomputphys{J. Comput. Phys} 
\def\mnras{MNRAS}
\def\pasj{PASJ}
\def\philtransrsoca{Phil. Trans. R. Soc. A} 
\def\solphys{Sol.~Phys.}

\hypersetup{draft}

\newcommand{\boltzmann}{k_\mathrm{B}}
\newcommand{\protonmass}{m_\mathrm{p}}

\title[Numerical simulations of the emerging plasma blob]{Numerical simulations of the emerging plasma blob into a solar coronal hole}

\author[Navarro et al]{
Anamar\'ia Navarro,$^1$ \thanks{E-mail:ana.navarro1@correo.uis.edu.co}
K. Murawski,$^2$      
D. W\'ojcik$^2$ and 
F. D. Lora-Clavijo$^1$ \\
$^{1} $ Grupo de Investigaci\'on en Relatividad y Gravitaci\'on, Escuela de f\'isica, 
Universidad Industrial de Santander,\\ A.A. 678, Bucaramanga, Colombia \\
$^2$ Group of Astrophysics, Institute of Physics, University of Maria Curie-Sk\l{}odowska, ul. Radziszewskiego 10,\\ PL-20-031 Lublin, Poland
}

\date{Accepted XXX. Received YYY; in original form ZZZ}
\pubyear{2018}

\begin{document}

\maketitle

\begin{abstract}

We numerically simulate emergence of a magnetic plasma blob into a solar coronal hole. This blob may be associated with granulation and therefore it has a weak magnetic field. Two-dimensional simulations are performed using the MAGNUS code which solves magnetohydrodynamic (MHD) equations, taking into account magnetic resistivity and thermal conduction. As a result of the interaction of the emerging blob with the ambient plasma, the magnetic lines experience reconnection with the blob getting flattened and deformed with time. Additionally, this process launches a vertical outflow of hot plasma and the chromosphere in its response increases its temperature.  We perform  parametric studies by varying the magnitude of the magnetic field of the blob and observing the net heating of the chromosphere. These studies are inspired by realistic simulations of granulation made with the use of two-fluid JOANNA code. In these simulations a number of magnetic blobs are detected in the convection zone and in the photosphere. From the numerical results, we conclude that as a result of granulation operating in a solar quiet region the emerging blob may trigger very complex dynamics in the upper regions of the solar atmosphere, and the associated outflows may be a source of heating of the chromosphere and possibly the solar corona.  

\end{abstract}

\begin{keywords}
MHD--waves--Sun: atmosphere
\end{keywords}

\section{Introduction}

Energy transport and dissipation through the solar atmospheric layers have been a subject of intensive investigations for a long time. These investigations are motivated by the aim to understand the energy supply for the high upper atmospheric layers. Magnetohydrodynamic (MHD) waves and flares are one of a few suitable candidates responsible for this \citep[e.g.][]{Arregui2015}. Numerous numerical models have been developed in order to study the contribution of different magnetic field configurations \citep[e.g.][]{DeMoortel2015}.  Particularly, emerging magnetic flux from the convection zone into the overlying atmosphere, modelled by twisted magnetic flux tubes which are susceptible to the Parker buoyancy instability, are associated with diverse solar phenomena like flare production, coronal mass ejections and jets \citep{2014LRSP...11....3C}. As a result of the advances in observational data and the constant improvement of numerical simulations, this active area of research in solar physics leads to determine diverse magnetic structures. See \cite{2011PASJ...63..407T}, \cite{Magara_2003} and references cited therein. The flux emergence has been mostly studied in active regions, where flux tubes are buffeted by strong convective magnetic fields. In quiet regions, flux emergence events have also been detected but at lower rates than in magnetic active regions. However, the physical processes seem to be similar in quiet and active regions \citep{Vargas_Gesztelyi_Bellot_2012}.

Since most of the studies carried out in this area have been developed under conditions of strong magnetic fields, in this work we explore the flux emerging possibly associated with granulation and therefore of a relatively weak magnetic field. On this basis, we propose a simple model to study the effects of a solitary magnetic blob launched from the base of the photosphere.
Our aim is to understand the conditions and effects of a weak magnetic flux emergence on the dynamics of the upper atmospheric layers. We consider a coronal hole being permeated by vertical ambient magnetic field and analyze the plasma dynamics resulting from flux emergence. Since the energy transfer is affected by dissipative mechanisms \citep[e.g.][]{Goodman_Kazeminezhad_2010}, we carry out our numerical simulations by taking into account a temperature dependent Spitzer resistivity for a fully ionized plasma and thermal conduction \citep{Spitzer1956}.

This paper is organized as follows. In Section \ref{sec_Joanna} we display magnetic blobs found in realistic numerical simulations of granulation performed with the use of two-fluid JOANNA code \citep{2018MNRAS.481..262W}. In Section \ref{sec_numerical_model} we introduce MHD equations. Section \ref{sec_results} presents the dynamics of the plasma. Additionally, we compare the changes of temperature in the chromosphere as a function of the magnetic field of the blob, by doing a parametric study. Finally, in Section \ref{sec_conclusions} we draw our conclusions.

\section{Magnetic blobs produced by granulation} \label{sec_Joanna}
\begin{figure*}
\centering
\includegraphics[width = 0.48\textwidth ]{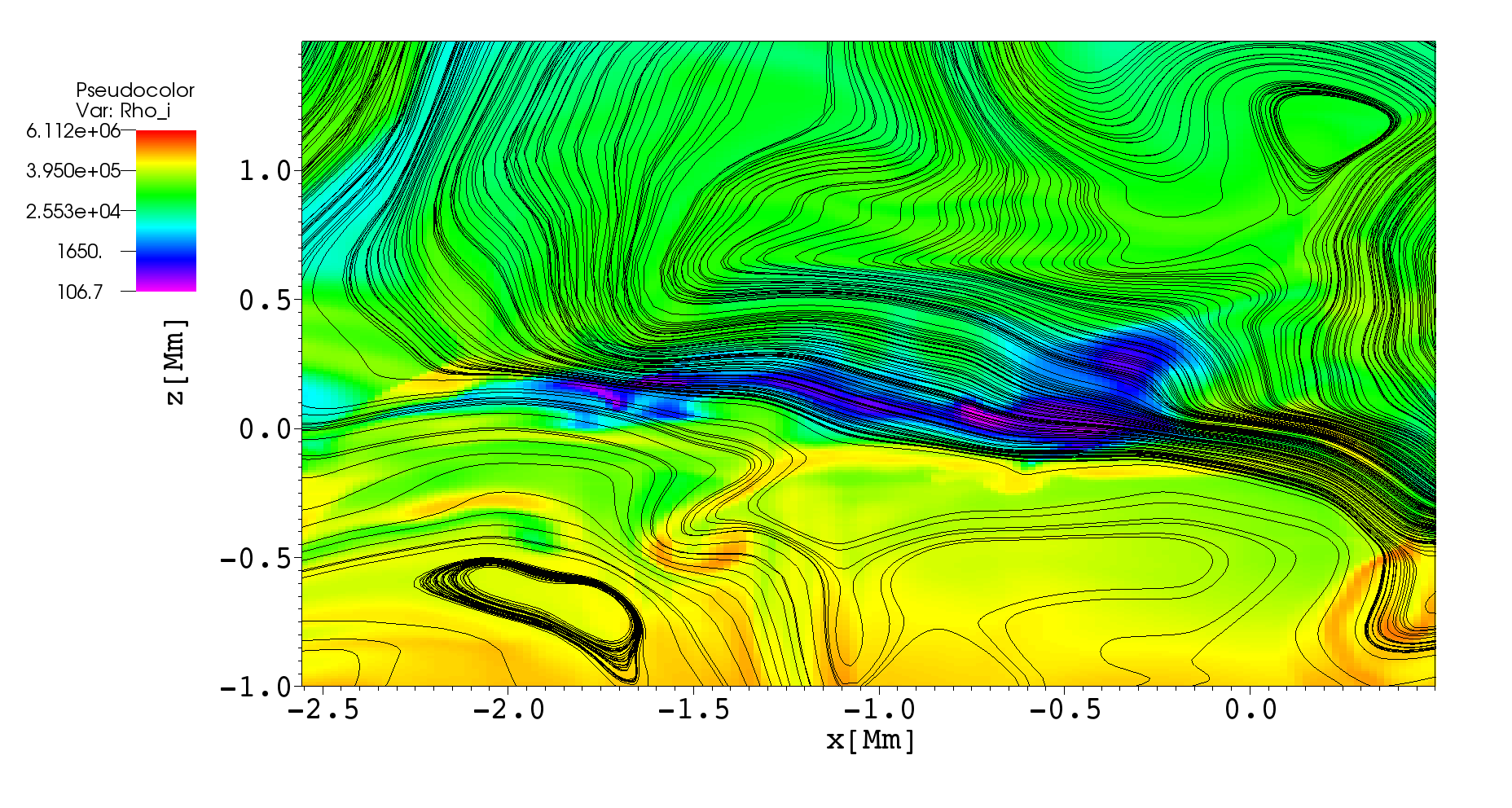}
\includegraphics[width = 0.48\textwidth ]{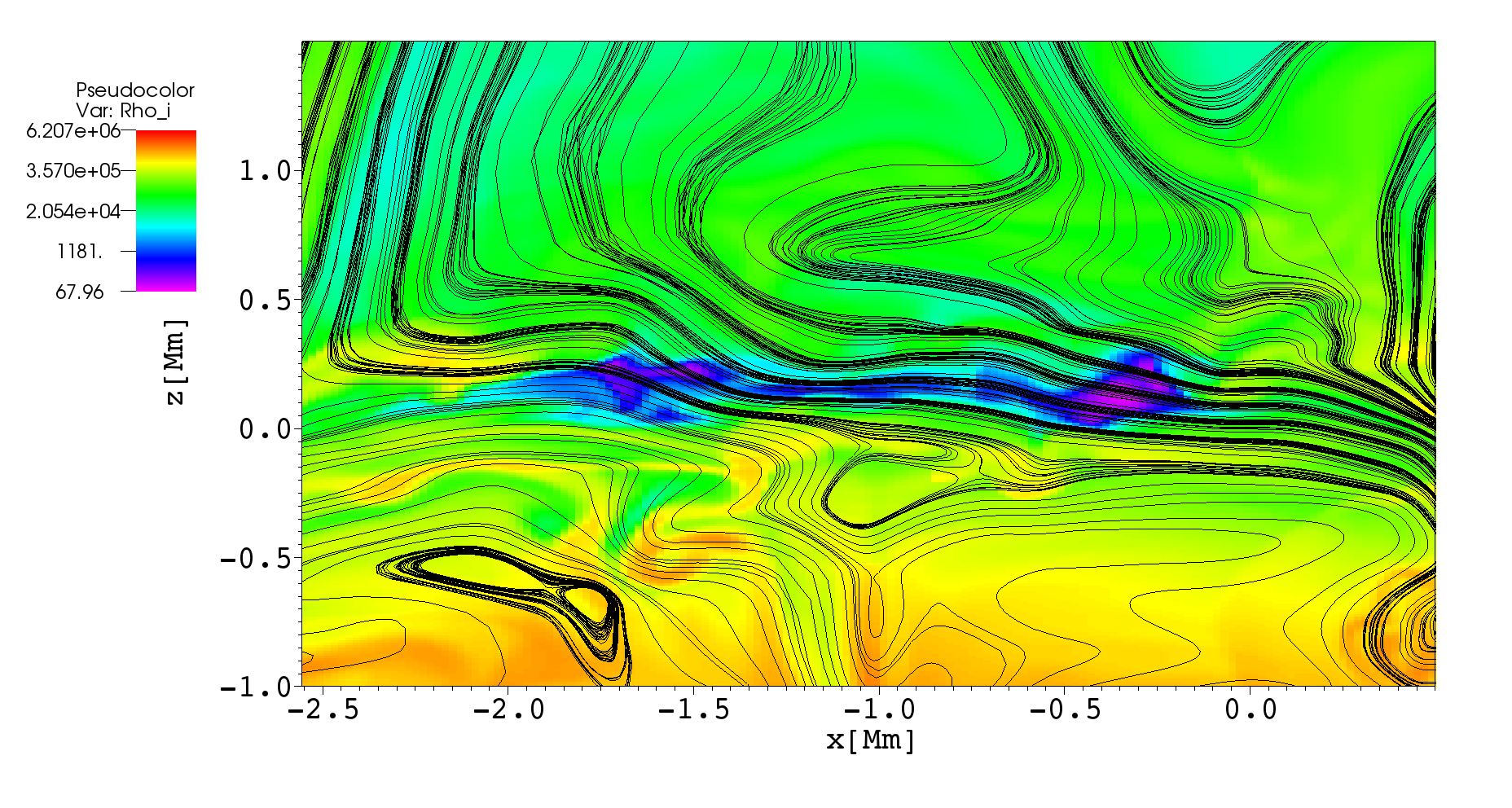}\\
\includegraphics[width = 0.48\textwidth ]{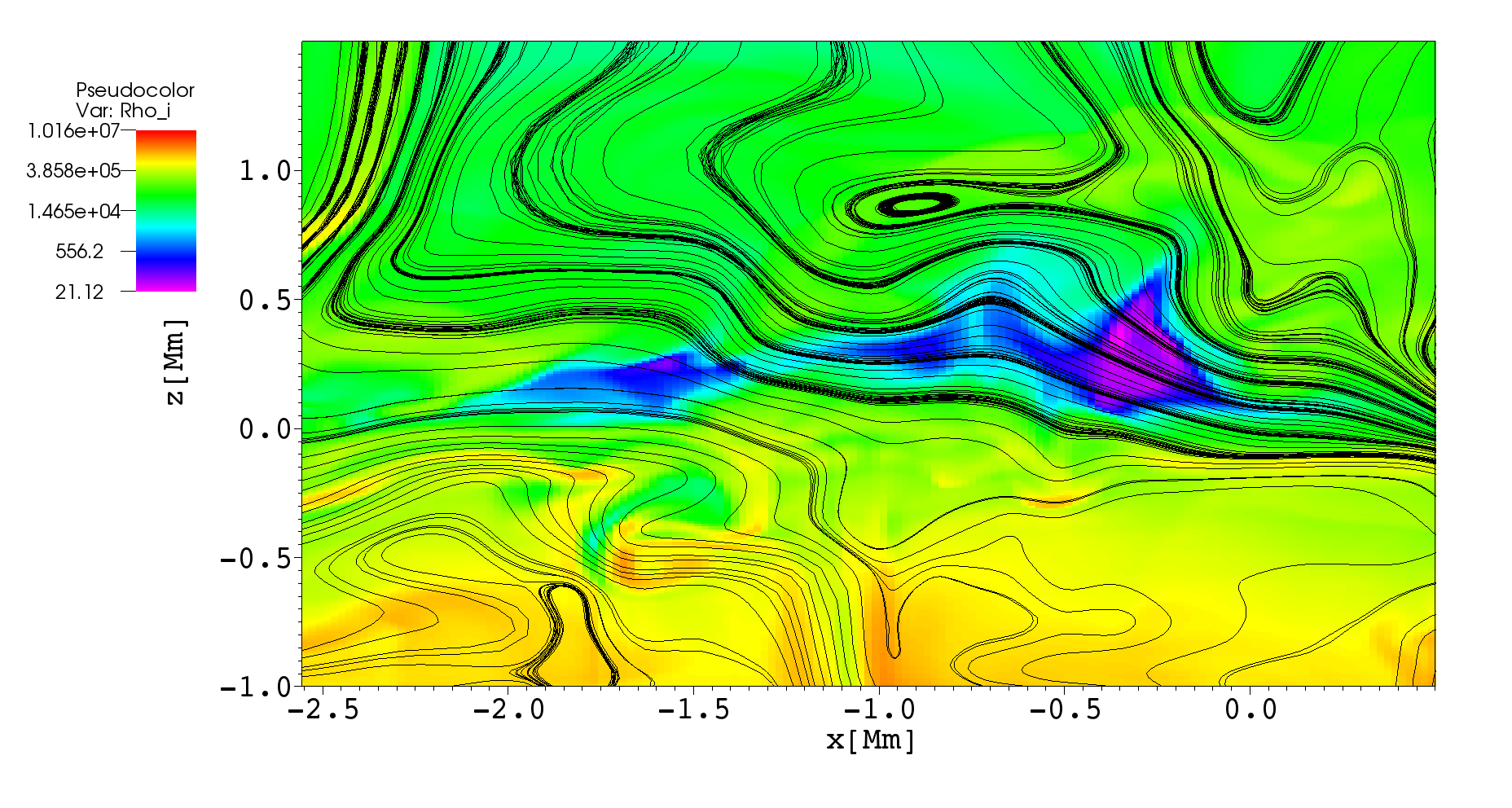}
\includegraphics[width = 0.48\textwidth ]{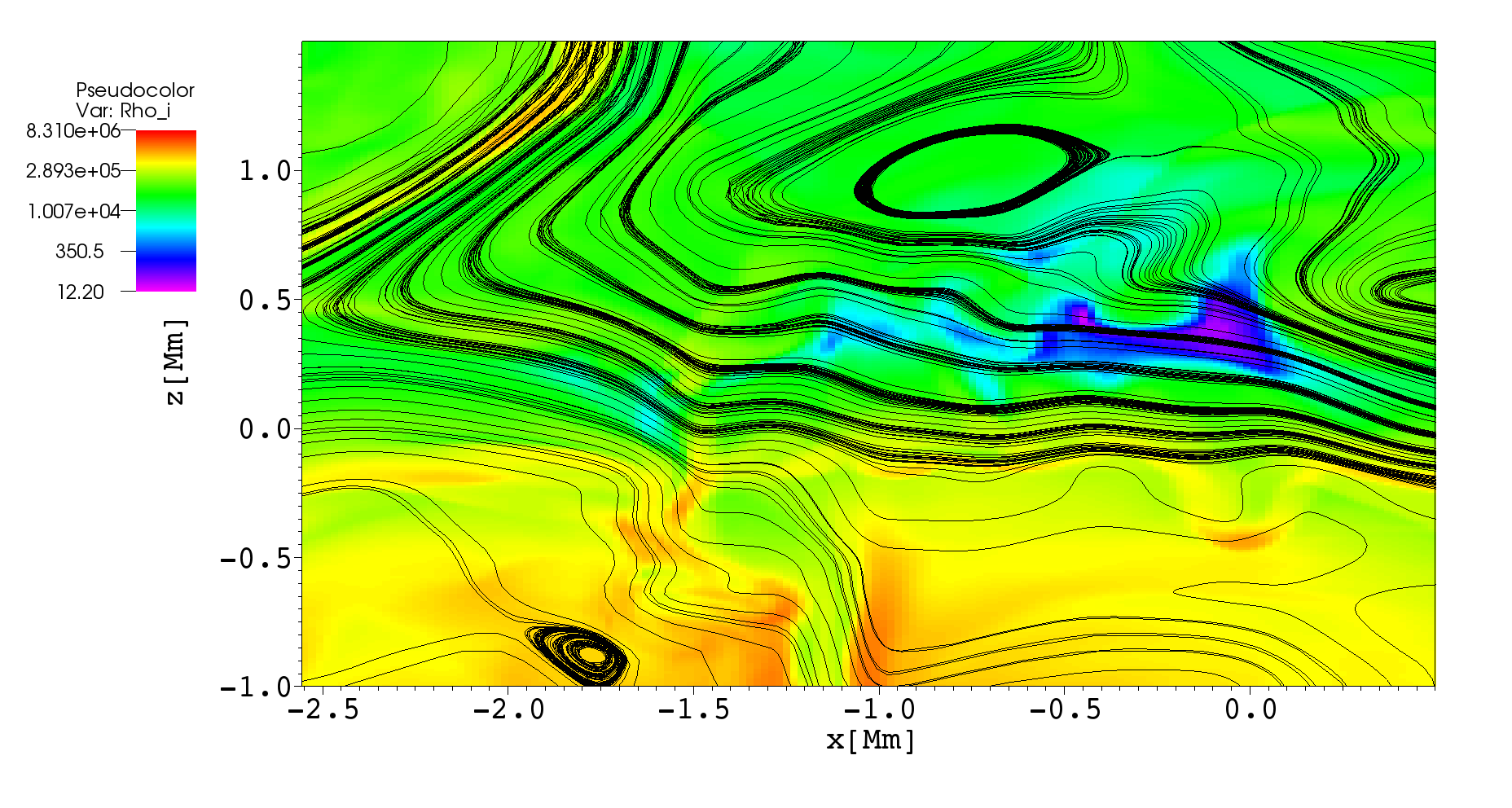}
\caption{Colormaps of the logarithm of ion mass density and magnetic field lines in the numerical simulations of granulation, using two-fluid JOANNA code at different consecutive times. \label{fig_granulation}}
\end{figure*}
Numerical simulations of solar granulation exhibit formation of magnetic blobs populating the convection zone and the photosphere. These simulations are performed with the use of two-fluid JOANNA code \citep{2018MNRAS.481..262W} which describes evolution of plasma consisting of two components, mainly of ionized (ions + electrons) fluid  and electrically neutral fluid (neutral atoms and molecules), and takes into account non-adiabatic terms such as thermal conduction, thermal convection and radiation. In these simulations the solar atmosphere is initially (at $t = 0$ s) permeated by a vertical magnetic field of strenght 5 G. In an advanced stage of the simulations, when granulation downdrafts were already produced, we found some interesting shapes of the magnetic field in the convection zone and in the photosphere. The convective movements in this zone resulted in bent magnetic field lines and in formation of blobs and other irregular shapes.  In Figure \ref{fig_granulation} we present colormaps of the logarithm of ion mass density and the magnetic field lines at four different times. Several blob structures can be spotted. The top-left panel shows an oval blob with its center located at about $[x = -1.9, z = -0.7]$ Mm, which splits into two blobs that move in different directions as the consecutive panels indicate. Four more blobs can be spotted in this figure, one with its center at about $[x = 0.3, z = 1.2]$ Mm, size of about 1 Mm in the top-left panel, a second one with its center at $[x = -1, z = -0.2]$ Mm and its size of 0.5 Mm (top-right panel), a third one with its center at $[x = -0.7, z = 0.8]$ Mm and its size of 0.5 Mm (bottom-left panel) and a fourth one with its center at $[x = -0.7,z = 1.0]$ Mm and its size of 1.5 Mm (bottom-right panel). These blobs exhibit magnetic field strengths from 9 G to 30 G, velocities from 3 km s$^{-1}$ to 30 km s$^{-1}$ and appear with a frequency of around 0.04 s$^{-1}$. They seem to be moving upwards, until they vanish or reconnect again with the surrounding magnetic field.

\section{Numerical model}\label{sec_numerical_model} 

We model the gravitationally stratified solar atmosphere with MHD equations, written in the conservative form as
\begin{eqnarray}
\partial_t \vec{U} + \partial_i \vec{F^i} = \vec{S} \, . \label{MHD_eq_ConservativeForm}
\end{eqnarray}
Here $\vec{U}$ is a state vector, $\vec{F^i}$ is the flux along $i-$axis and $\vec{S}$ is a source vector, all are given by
\begin{eqnarray}
& & \vec{U} = \left[ \begin{array}{c}
\varrho \\
\varrho \vec{v} \\
 E \\
\vec{B} 
\end{array} 
  \right] \, ,  \hspace{3mm}  
\vec{F^i} = \left[ \begin{array}{c} \varrho v^i \\  
 \varrho v^i v_j - \frac{B^i B_j}{\mu_\mathrm{0}} + p_\mathrm{T}\delta^i_j \\
( E + p_\mathrm{T})v^i - \frac{B^i(\vec{B}\cdot \vec{v})}{\mu_\mathrm{0}} \\
v^i B_k - v_k B^i   \end{array}  \right] \, , \qquad   \end{eqnarray} \begin{eqnarray}
 \vec{S} = \left[ \begin{array}{c}
0 \\
- \varrho \vec{g} \\ 
-\varrho \vec{v}\cdot \vec{g} - \nabla \cdot \left( \vec{q} + \frac{\eta}{\mu_\mathrm{0}} \vec{J} \times \vec{B} \right) \\
-\nabla \times \eta \vec{J}
 \end{array} \right] \, .
\end{eqnarray}
The symbol $\varrho$ denotes mass density, $\vec{v}$ the velocity, $\vec{B}$ the magnetic field, $p$ the gas pressure, $\mu_\mathrm{0}$ the magnetic permeability of free space, $\vec{J} =  \nabla \times \vec{B}/\mu_\mathrm{0}$ is the current density,  $p_\mathrm{T} = p + B^2/2\mu_\mathrm{0}$ the total pressure,  $\eta$ the magnetic resistivity coefficient, $\vec{q}$ the heat flux, $\vec{g} = [0,0,-g]$ the gravitational acceleration with its magnitude of 274 m s$^{-2}$, and  
\begin{eqnarray}
E = \frac{\varrho v^2}{2} + \frac{p}{\Gamma-1} + \frac{B^2}{2 \mu_0} \, , 
\end{eqnarray}
is the total energy density. The fluid obeys the ideal gas law with adiabatic index $\Gamma = 5/3$.

The thermal conduction operates along magnetic field lines by means of the classical model for magnetized plasma \citep{Spitzer2006}, leading to
\begin{eqnarray}
\vec{q} =  \kappa T^{5/2} ( \vec{B} \cdot \nabla T ) \vec{B}/B^2 \, , 
\end{eqnarray}
where $\nabla T$ is the temperature gradient and the thermal conductivity coefficient $\kappa = 10^{-11}$ W m$^{-1}$ K$^{-7/2}$. The implemented resistivity coefficient, $\eta$, is the Spitzer's model for a fully ionized plasma, given in SI units by
\begin{eqnarray}
\eta = \frac{65.359 \ln \Lambda }{\ T^{3/2}} \, , \label{eq_eta}
\end{eqnarray}
where $\ln \Lambda $ is the Coulomb logarithm, which is set to 20 for the performed simulations.

\subsection{Initial hydrostatic state} \label{subsec_initial_data} 

To obtain the initial state of mass density and gas pressure, we first set $\partial_t = 0$ and $\vec{v} = 0$ in equation (\ref{MHD_eq_ConservativeForm}), and get the hydrostatic equilibrium equation as
\begin{eqnarray}
\varrho g + \frac{\partial p}{\partial z} = 0 \ .\label{eq_equilibrium}
\end{eqnarray}
From the ideal gas law, the gas pressure is given by 
\begin{eqnarray}
p =  \frac{\boltzmann}{\protonmass} \varrho T \, , \label{eq_EOS}
\end{eqnarray}
where $\protonmass$ is the proton mass, $\boltzmann$ is the Boltzmann's constant and $T$ is the temperature. From these two equations, the hydrostatic mass density and gas pressure are derived as functions of a temperature profile $T(z)$ in the following form:  
\begin{eqnarray}
& & p(z) =  p_\mathrm{0} \exp{\left( -\frac{\protonmass g}{\boltzmann} \int_{z_\mathrm{0}}^z \frac{\mathrm{d}\tilde{z}}{T(\tilde{z})}  \right)} \, , \qquad  \\
& & \varrho(z) = \frac{\protonmass}{\boltzmann} \frac{p(z)}{T(z)}  \, ,
\end{eqnarray}
where $z_\mathrm{0} = 10$ Mm is the reference height.
In our case, we use the semiempirical model developed by \cite{2008AvrettLoeser}.

\begin{figure*}
\centering
\includegraphics[width = 0.7\textwidth]{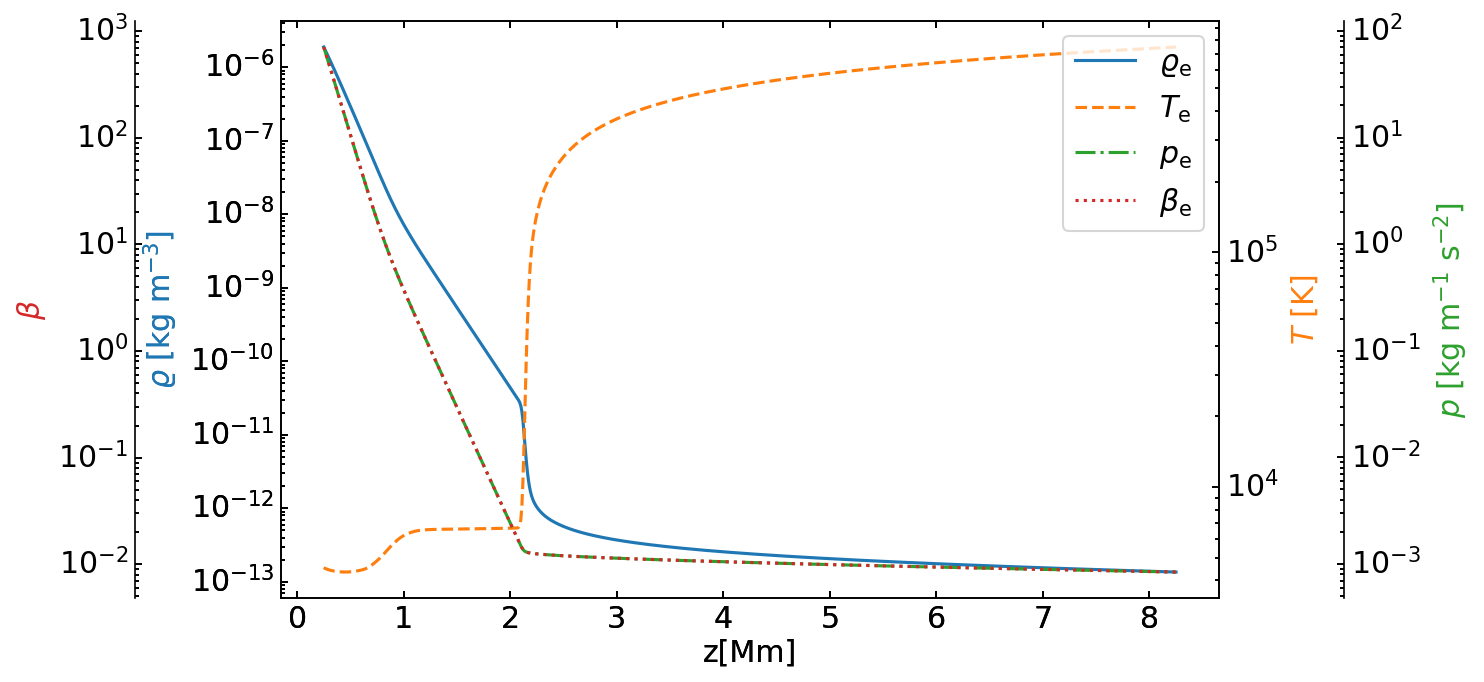} 
\caption{Hydrostatic profiles of the plasma$-\beta = 2\mu_0 p/B^2$, the mass density $\varrho$, the temperature $T$ and the gas pressure $p$ versus height. \label{initial_profiles}}
\end{figure*}

The initial hydrostatic equilibrium state for the simulations is supplemented by a uniform vertical magnetic field, $\vec{B} = [0, 0, B]$, of its magnitude $B = 5$ G. Figure  \ref{initial_profiles} displays the plasma$-\beta = 2\mu_0 p/B^2$, the mass density $\varrho$, the temperature $T$ and the gas pressure $p$ versus height $z$. The temperature profile starts from 
$4.5\times 10^3$ K  at $z = 0.25 $ Mm,
 decreases to a minimum of $4.3\times 10^3$ K at $z = 0.55$ Mm, rises to 
 $6.8\times 10^3$ K at the high chromosphere ($z = 2.1$ Mm) and then grows abruptly in the transition region and smoothly increases to 
 $6.5\times 10^5$ K at $z = 6.25$ Mm. Accordingly, the mass density attains a value of $1.88\times 10^{-6}$ kg m$^{-3}$ at $z = 0.25$ Mm, decreases exponentially to $2.8\times 10^{-11}$ kg m$^{-3}$ in the top chromosphere and then experiences a sudden fall-off below the transition region, and smothly declines to $1.7\times 10^{-13}$ kg m$^{-3}$ at $z = 6.25$ Mm. The gas pressure reaches a value of 70.6 kg m$^{-1}$ s$^{-2}$ at $z = 0.25$ Mm, subsides to 1.5$\times 10^{-3}$ kg m$^{-1}$ s$^{-2}$ at $z = 2.1$ Mm, falls off rapidly in the transition region and then smothly reaches 9.2$\times 10^{-4}$  m$^{-1}$ s$^{-2}$ at $z = 6.25$ Mm. The plasma-$\beta$ starts from about $710$ at $z = 0.25$ Mm, reaches a value of 1 at $z = 1.2$ Mm and 0.009 at $z = 6.25$ Mm. 

\subsection{Simulation box and boundary conditions}\label{subsec_simulationbox} 

We carry on numerical simulations with MAGNUS code \citep{magnus}, within a simulation box covered by a uniform grid of 5 km resolution in both $x-$ and $z-$directions with the extensions $[-2.0, 2.0]\times[0.25, 8.25]$ Mm$^2$. The Courant-Friedrichs-Lewy (CFL) number is set equal to 0.1. The used numerical methods are the HLLE Riemann solver \citep{Harten_etal_1983}, the van Leer slope limiter \citep{Vleer1977}, and a third order total variation diminishing Runge-Kutta \citep{Abramowitz1972}. To prevent the growth in time of divergence of the magnetic field we adopt the Flux Constrained Transport method \citep{CT_evans, CT_Balsara}. The MAGNUS code solves the one-fluid MHD equations and take into account resistivity and thermal conduction (thermal convection is not included) in contrast to the ones of the Joanna code that is more realistic since it solves the two fluid equations, taking into account thermal conduction, thermal convection and radiation.

The implemented boundary conditions are outflow at the lateral sides and fixed to the hydrostatic plasma quantities at the bottom and top sides. The bottom boundary is supplemented by the magnetic field which models the magnetic flux emergence according to
\begin{eqnarray}
  B_x(x,z,t) &=& - \frac{B_\mathrm{0} [z-z_\mathrm{c}(t)]}{\omega} \exp \left( -\left\lbrace x^2 + [z - z_\mathrm{c}(t)]^2 \right\rbrace / \omega^2 \right)  \, ,  \\
  B_z(x,z,t) &=&  \frac{B_\mathrm{0} x }{\omega} \exp\left(-\left\lbrace x^2 + [z - z_\mathrm{c}(t) ]^2 \right\rbrace / \omega^2 \right)  \, .
\end{eqnarray}
This flux mimics a circular magnetic blob with its center, located initially at $x = 0$, $z = - 1.5 \  \omega $, with $\omega$ being the blob's width. The blob moves upwards with a constant speed $v_\mathrm{b} = 5$ km s$^{-1}$ and its vertical location is given by 
\begin{eqnarray}
z_\mathrm{c}(t) =  - 1.5 \  \omega + v_\mathrm{b} \ t \, .
\end{eqnarray}
 Here we set $\omega = 200$ km  and $B_\mathrm{0}$ is 180 G (unless otherwise stated) which corresponds to a maximum value of blob's magnetic field of 77 G. The selection of these values is in accordance with the detected blobs of two-fluid simulations of granulation described in Section \ref{sec_Joanna}.

\section{Numerical Results}\label{sec_results}

\subsection{Morphology of the blob emergence} 

\begin{figure}
\centering
\includegraphics[width = 0.45\textwidth ]{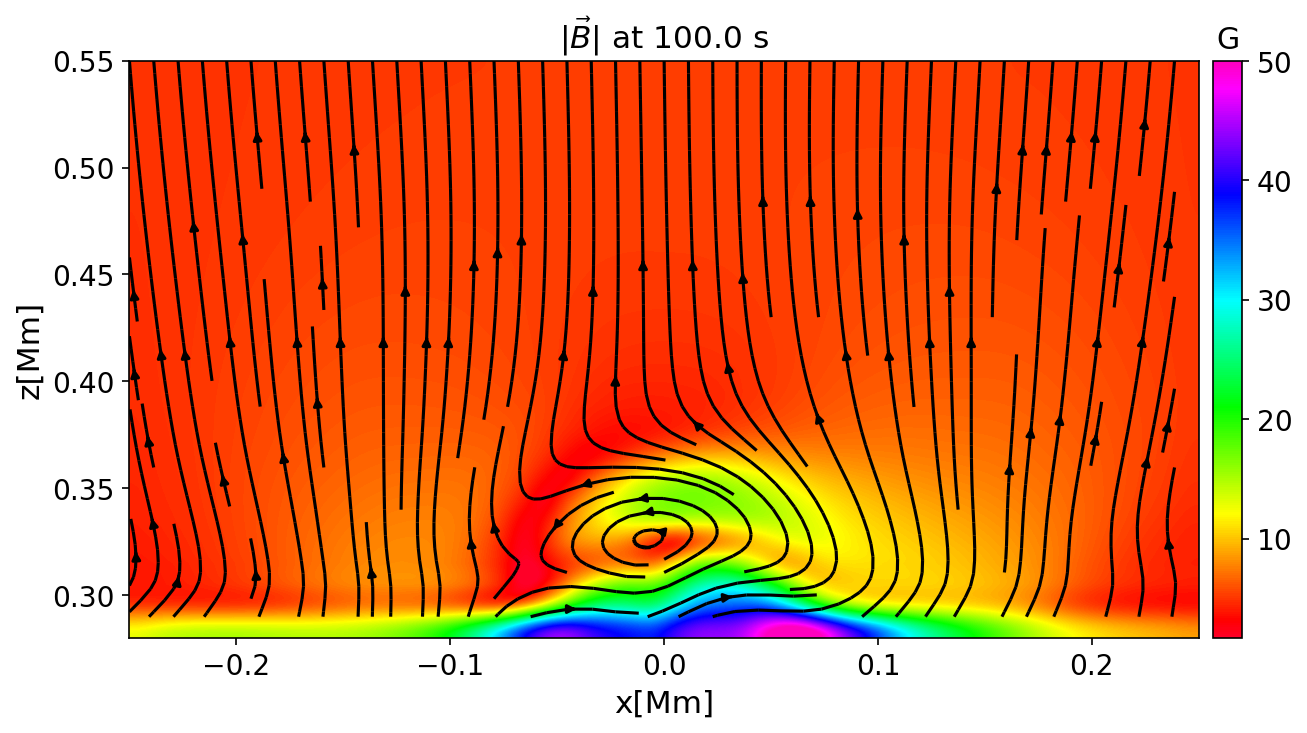}\\
\includegraphics[width = 0.45\textwidth]{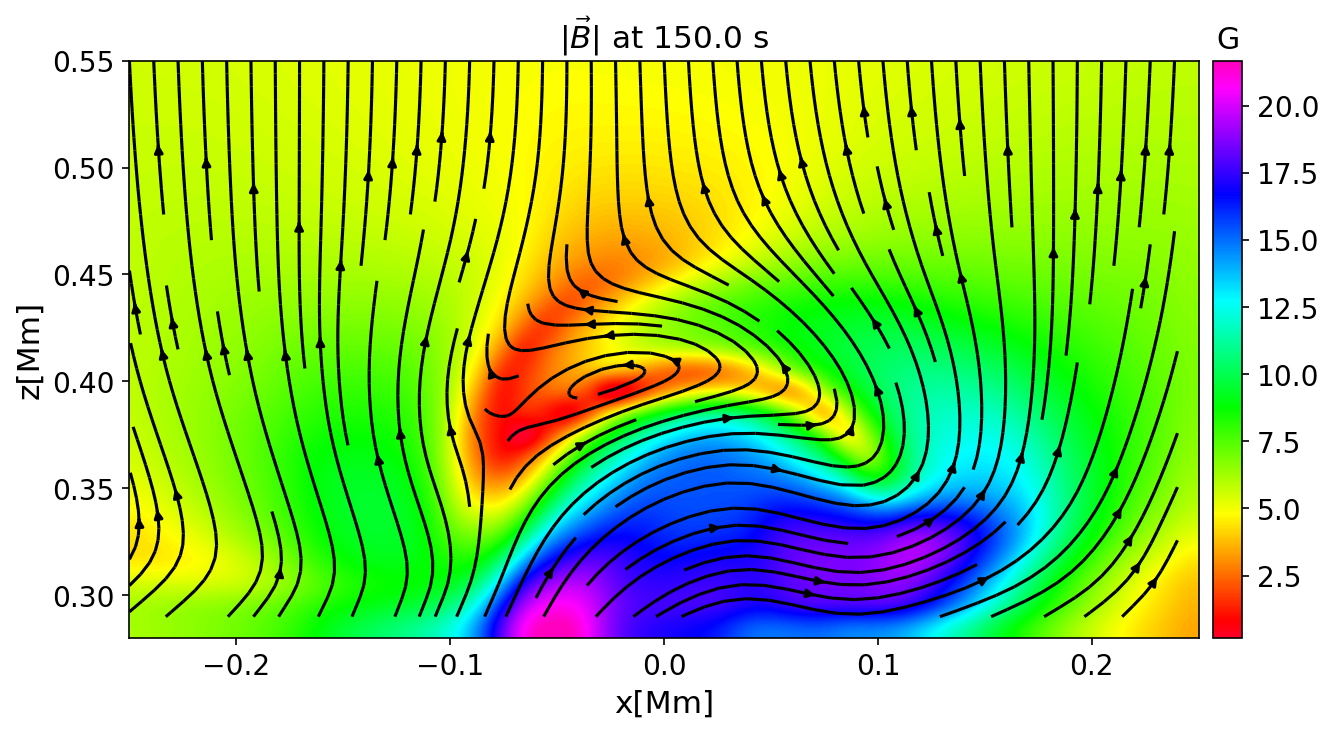}\\
\includegraphics[width = 0.45\textwidth]{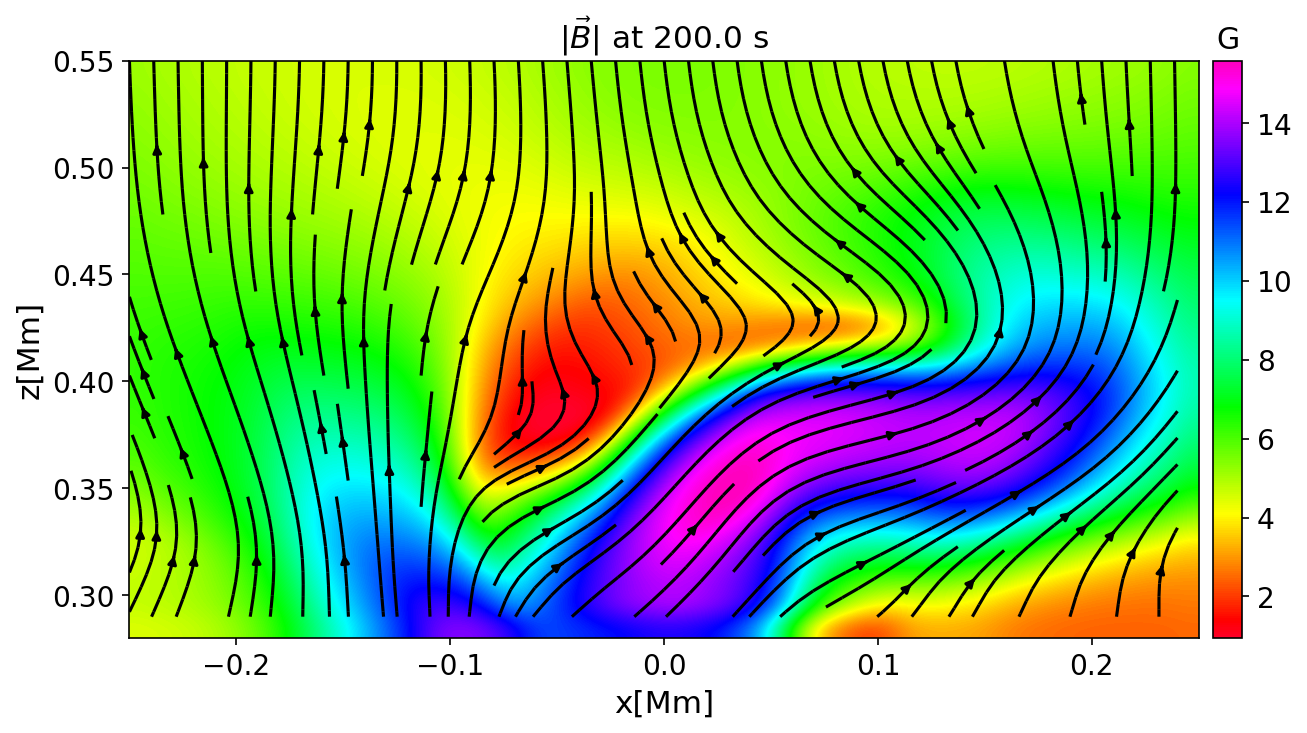}
\caption{Colormaps of magnetic field magnitude and its lines at $t = 100$ s (top), $t = 150$ s (middle), and $t = 200$ s (bottom). \label{magB_linesB} }
\end{figure}

The emerging magnetic blob modifies the solar atmosphere in various ways. Figure \ref{magB_linesB} displays the colormaps of the magnetic field magnitude and the magnetic field lines at three different times. At $t = 100 $ s, a new blob is produced as a result of magnetic reconnection with the ambient field; by this time almost all the emerging flux has entered the atmosphere. The top panel shows that only half of the blob was able to reconnect since the new blob has a radius of 100 km, which corresponds to half of the initial size of the emerging blob. The middle panel shows that at a later time; this blob has been continued travelling upwards and its shape has been flattened. Finally, the right panel displays that at $t = 200 $ s the blob has vanished completely. This kind of behaviour is in agreement with the observed dynamics of the blobs generated by granulation in the two-fluid simulations described in Section \ref{sec_Joanna}. 

\begin{figure*}
\centering
\includegraphics[height = 0.25\textheight]{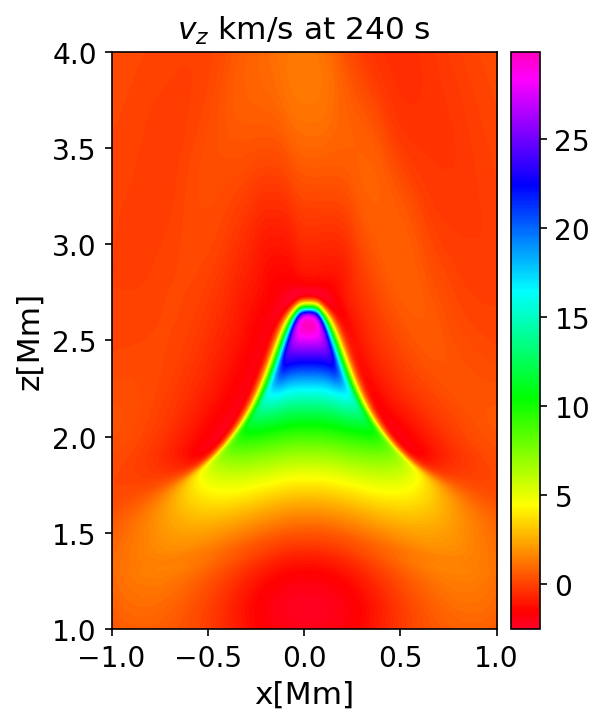}
\includegraphics[height = 0.25\textheight]{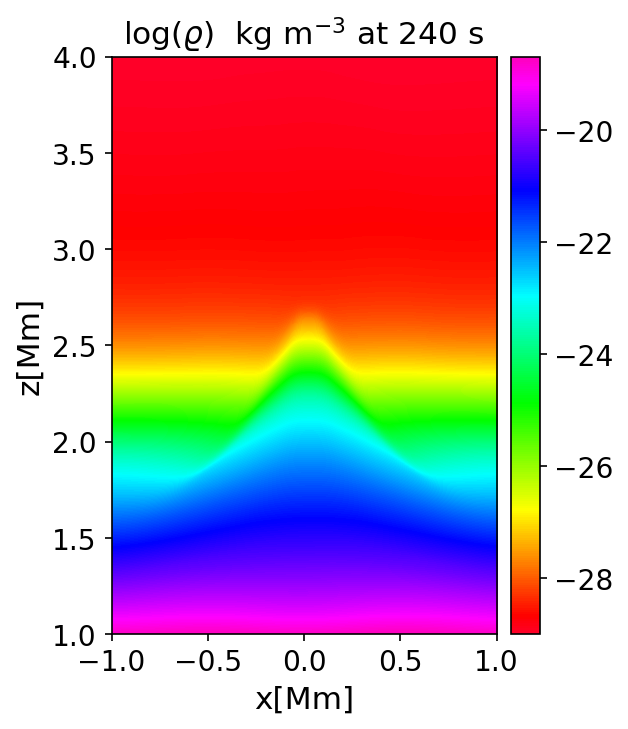}\\
\includegraphics[height = 0.25\textheight]{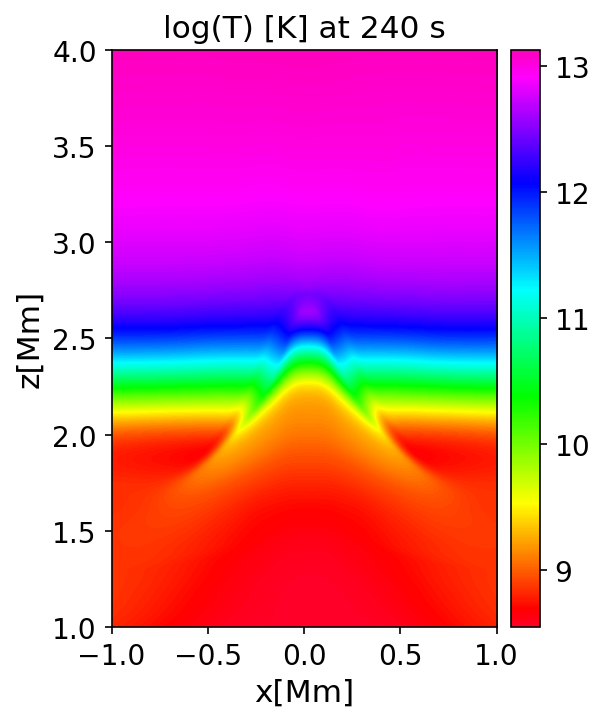}
\includegraphics[height = 0.25\textheight]{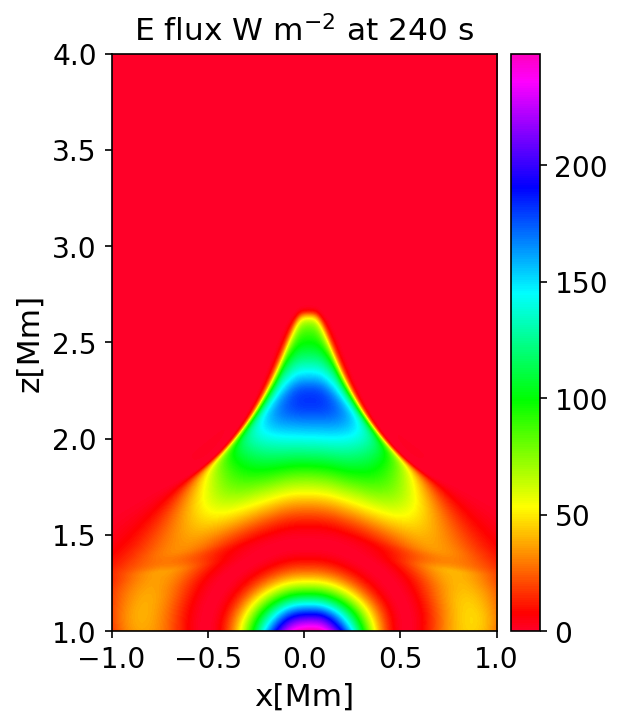}
\caption{Colormaps of vertical velocity (top-left), logarithm of mass density (top-right), logarithm of temperature (bottom-left), energy flux (bottom-right) at $t = 240$ s. \label{Fig_vz_rho_T} }
\end{figure*}

Figure \ref{Fig_vz_rho_T} shows the colormaps of vertical velocity, logarithm of mass density, logarithm of temperature and vertical energy flux in the upper region at time $t = 240$ s. From the emergence of the blob, a shock wave is triggered into the higher layers. At $t = 240$ s its front has almost reached the base of the corona and has a speed of 30 km s$^{-1}$. Accordingly, the colormap of the density logarithm shows an outflow of mass traveling with the shock wave in a spicule shape. The plot of temperature indicates that this moving plasma has a larger temperature than its surroundings. The vertical energy flux $\varrho v_z^2 \sqrt{\Gamma p/\varrho}$ has a value of 300 W m$^{-2}$ in the low chromosphere, and  200 W m$^{-2}$ in the shock front. The latter is equal to 20$\%$ of the estimated energy flux loss in a coronal hole \citep{Withbroe_Noyes_1977}. It is worth mentioning that the emerging magnetic flux in the way that it is introduced by the boundary conditions may generate the observed shock by the enhanced total pressure, since the force balance is not satisfied. In addition, a part of the increase of the temperature may be related to this effect.

\begin{figure}
\centering
\includegraphics[width = 0.45\textwidth]{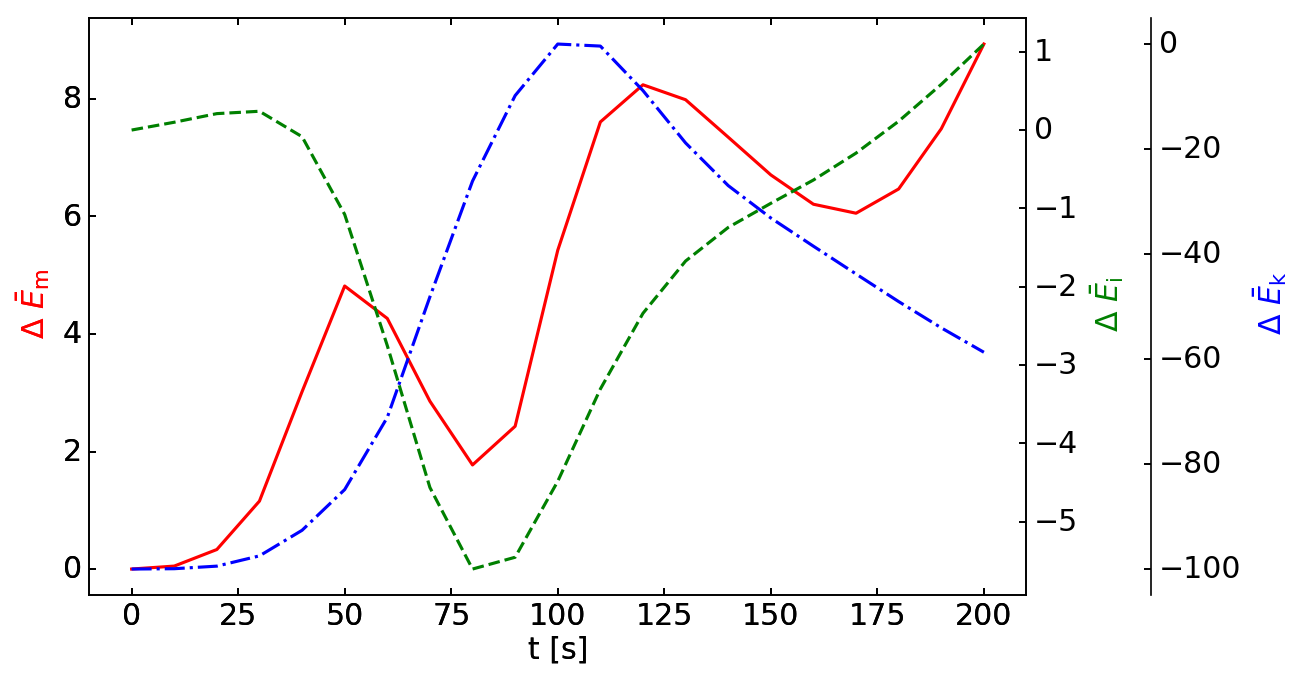}\\
\caption{Variations of the magnetic (first axis, solid line), internal (second axis, dashed line) and kinetic (third axis, dash-dotted line) energy as a function of time. \label{fig_E_changes} }
\end{figure}

The magnetic, kinetic and internal energies of the plasma, averaged in the numerical domain [$x_1 = -1.0$, $x_2 = 1.0$]$\times$[$z_1 = 0.25$, $z_2 = 2.1$] Mm$^2$, are evaluated as
\begin{eqnarray}
 \bar{E}(t) &=& \frac{1}{A} \int_{x_\mathrm{1}}^{x_\mathrm{2}} \int_{z_\mathrm{1}}^{z_\mathrm{2}} E(x,z,t)  \ \mathrm{d}x \ \mathrm{d}z \, , \label{E_integral} 
\end{eqnarray}
where $E(x,z,t)$ is an energy density, such as the magnetic $E_\mathrm{m} = B^2/2\mu_0$, the kinetic $E_\mathrm{k} = \rho v^2/2$, and the internal $E_\mathrm{i} = p/(\Gamma-1)$ and $A = (x_2-x_1)(z_2-z_1)$ is the area of the region of interest. The relative change of these quantities with respect to the initial state is 
\begin{eqnarray}
 \Delta \bar{E} = \frac{ \left( \bar{E}(t) - \bar{E}_\mathrm{e}\right) }{ \bar{E}_\mathrm{e} }  \times 100 \% \, ,  \label{Delta_E}
\end{eqnarray} 
where $E_\mathrm{e}$ stands for the value in equilibrium. However, since the initial value of the kinetic energy is zero, we use as a reference its value at $t = $ 100 s. Figure  \ref{fig_E_changes} displays variations of the energies versus time, the first axis corresponds to the variations of magnetic energy (solid line). The magnetic energy increases as the blob emerges until $t = 50 $ s, where the center of the blob enters the domain. Since this center has lower magnetic field there is a local minimum in the change of the energy at $t = 80 $ s. After this moment, the energy starts to increase again until $t = 125 $ s, at which the blob is formed completely and  reaches its highest point. From this moment, the blob starts to fall, gets flattened and deformed until it vanishes, causing a decrement on the magnetic energy. The second axis gives the changes of the internal energy density; in the early stages of the blob emergence it increases, and starts to decrease from $t = 30 $ s until it reaches its minimum value at $t = 80 $ s, which is the same time as the magnetic energy does. After this moment of time, as the magnetic energy grows, the internal energy starts to increase linearly, but after $t = 125 $ s, when the blob starts to vanish, its growth rate decreases. The kinetic energy grows in a gaussian way, attains its maximum value at $t = 100$ s, when the blob emerges completely into the domain and after this moment energy starts to decrease; at this moment the shock wave leaves the chromosphere and it is travelling into the above layers. 

\begin{figure}
\centering
\includegraphics[width = 0.45\textwidth]{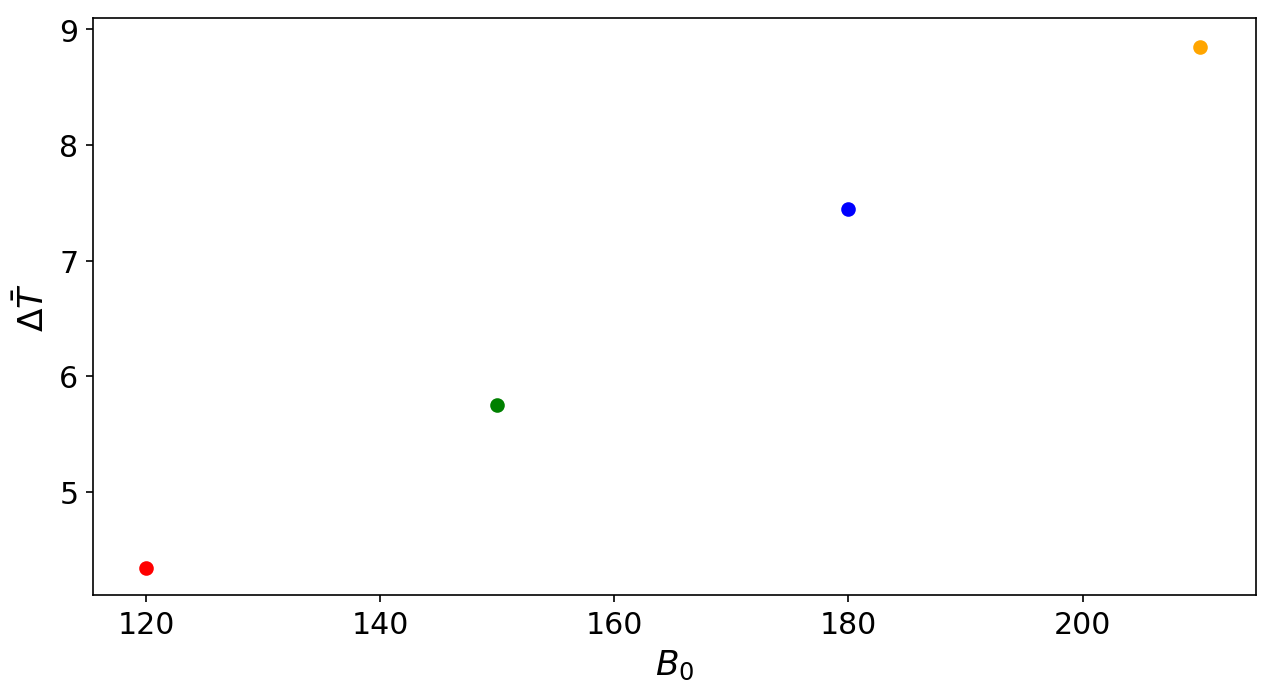}\\
\caption{Relative temperature variation in the chromosphere, in the region [$x_1 = -1.0$, $x_2 = 1.0$]$\times$[$z_1 = 0.25$, $z_2 = 2.1$] Mm$^2$ at $t=210$ s as a function of $B_0$, the parameter associated to the magnitude of the magnetic field of the blob.  \label{Fig_comparisons} }
\end{figure}

We perform parametric studies by varying the magnitude of the magnetic field of the blob $B_0$. In Figure \ref{Fig_comparisons} we present spatially averaged relative variations change in plasma temperature, $\Delta \bar{T}$, in the chromosphere evaluated at $t = 210 $ s as a function of $B_0$. These relative variations were calculated in the same way as the changes in energy in equation (\ref{Delta_E}). From the figure we infer that temperature variation $\Delta\bar{T}$ grows close to linear trend with the magnetic field of the blob. In the case of $B_0 = 180 $ G, that corresponds to the simulation described throughout the text, $\Delta \bar{T}$ is about 7.5$\%$.

\section{Discussion and conclusions}
\label{sec_conclusions}

In this paper, we modelled the emergence of a magnetic blob ejected from the bottom of the photosphere into a coronal hole. We described the morphology of magnetic field lines, vertical velocity, mass density, temperature, energy flux and net change in temperature of the chromosphere. We found that the plasma dynamics developed by the emerging blob is very interesting since it produces some outflows of hot plasma into the higher layers, the vertical energy flux is considerably high and the dissipative mechanisms cause heating of the chromospheric plasma.  We believe that since there exists an evidence of magnetic blobs formation in the convection zone and the photosphere, the convective flows may push magnetic blobs up to the photosphere and studies of the effects of a single solitary blob is crucial in order to understand the scope of the effects associated with it. However we would like to point out that the mechanism of heating through emerging blobs still remains unclear but nonetheless the reported results show that they may be effective as a heating mechanism since the occurrence frequency found is not negligible.

\section*{Acknowledgments}
The authors express their thanks to the referee for his/her comments.  A. N. thanks to the financial support from COLCIENCIAS, Colombia, under the program ``Becas Doctorados Nacionales 647'' and Universidad Industrial de Santander to cover her research stay at University of Maria Curie-Sk\l{}odowska, Lublin, Poland.  F.D.L-C gratefully acknowledges the financial support from VIE-UIS, grant number 2314 and by COLCIENCIAS, Colombia, under Grant No. 8863. K.M.'s and D.W.'s work was done within the framework of the project from the
Polish Science Center (NCN) Grant Nos. 2017/25/B/ST9/00506 and 2017/27/N/ST9/01798." The JOANNA code was developed by Darek W\'ojcik.


\end{document}